\documentclass[12pt]{iopart}
\usepackage{graphicx}

\begin{document}

\title{Dark energy cosmology with generalized linear equation of state}

\author{E. Babichev\dag,
V. Dokuchaev\dag\footnote[3]{dokuchaev@inr.npd.ac.ru}
and Yu. Eroshenko\dag }

\address{\dag\ Institute for Nuclear Research of the Russian Academy
of Sciences \\
60th October Anniversary Prospect 7a, 117312 Moscow, Russia}

\begin{abstract}
Dark energy with the usually used equation of state $p=w\rho$,
where $w=const<0$ is hydrodynamically unstable. To overcome this
drawback we consider the cosmology of a perfect fluid with a
linear equation of state of a more general form
$p=\alpha(\rho-\rho_0)$, where the constants $\alpha$ and $\rho_0$
are free parameters. This non-homogeneous linear equation of state
provides the description of both hydrodynamically stable
($\alpha>0$) and unstable ($\alpha<0$) fluids. In particular, the
considered cosmological model describes the hydrodynamically
stable dark (and phantom) energy. The possible types of
cosmological scenarios in this model are determined and classified
in terms of attractors and unstable points by the using of phase
trajectories analysis. For the dark energy case there are possible
some distinctive types of cosmological scenarios: (i) the universe
with the de Sitter attractor at late times, (ii) the bouncing
universe, (iii) the universe with the Big Rip and with the
anti-Big Rip. In the framework of a linear equation of state the
universe filled with an phantom energy, $w<-1$, may
have either the de Sitter attractor or the Big Rip.

\end{abstract}

\maketitle

\section{Introduction}
The astronomical observations indicate that the expansion of our
Universe accelerates \cite{acceler}. In the framework of the
General Relativity this means that about two thirds of the total
energy density of the Universe consists of dark energy: the still
unknown component with a relativistic negative pressure
$p<-\rho/3$. The simplest candidate for dark energy is the
cosmological $\Lambda$-term or vacuum energy. During the
cosmological evolution the $\Lambda$-term component has the
constant (Lorentz invariant) energy density $\rho$ and pressure
$p=-\rho$. However the $\Lambda$-term requires that the vacuum
energy density be fine tuned to have the observed very low value.
For this reason the different forms of dynamically changing dark
energy with an effective equation of state $w\equiv p/\rho<-1/3$
were proposed instead of the constant vacuum energy density. As a
particular example of dark energy, the scalar field with some a
slow rolling potential (quintessence) \cite{CaDaSt} is often
considered. The possible generalization of quintessence is a
$k$-essence \cite{ArMuSt}, the scalar field with a non-canonical
Lagrangian.

Present observation data constrain the range of equation-of-state
of a dark energy as $-1.38<w<-0.82$ \cite{w}. Recently the Chandra
observations \cite{Chandra} gave the similar restrictions on the
value of $w$. These data do not exclude the possibility of our
universe to be filled with a phantom energy \cite{Caldw}: the
energy with a super-negative equation of state $w<-1$ (note,
however, that the dark energy with the equation of state $w$
evolving from the quintessence-like $w>-1$ in past, to
phantom-like $w<-1$ at present, provides the best fit for
supernova data \cite{cross}). The different aspects of phantom
cosmology were considered in \cite{phcosm}. The phantom energy is
usually associated with the phantom or ghost fields --- the scalar
fields with wrong-sign kinetic term. It is known that such fields
are unstable due to quantum instability of vacuum, unless the
kinetic term stabilized by the ultra-violet cutoff \cite{Cline}.
The intriguing possibility of constructing of effective $w<-1$
from the scalar field with correct-sign kinetic term was proposed
in \cite{Onemli}. In \cite{SahSht} the authors constructed the
theory with the effective $w<-1$ in the braneworld models.

The presence of phantom energy in the universe leads to the
interesting physical phenomena: the possibility of the Big Rip
scenario \cite{Caldw}, the black hole mass decreasing by phantom
energy accretion \cite{BDE} and a new type of wormhole evolution
\cite{GonzDiaz}. The alternatives to a scalar field model are the
perfect fluid models such as a Chaplygin gas \cite{Chaplygin}. Hao
and Li \cite{HaoLi} demonstrated that $w=-1$ state is an attractor
for the Chaplygin gas and equation of state of this gas could
approach to this attractor from the either $w<-1$ or $w>-1$ sides.

In this paper we analyze the perfect fluid model with a general
linear equation of state $p=\alpha(\rho-\rho_0)$ where $\alpha$
and $\rho_0$ are constants. This is a generalization of a
homogeneous linear equation of state ($\rho_0=0$) and is suitable
for the modelling either the linear gas with $p>0$ or the dark
energy with $p<0$. The advantage of the use of the general linear
equation of state is the possibility to describe the dark energy
with a positive squared sound speed (for the usually considered
equation of state $p=w\rho$ the dark energy is hydrodynamically
unstable, because $\partial p/\partial\rho=w<0$). The considered
linear model is reduced to the perfect fluid model with $w=const$
at the particular case of $\rho_0=0$.

In the framework of this model it is possible to find analytical
cosmological solutions for the arbitrary values of parameters
$\alpha$ and $\rho_0$. We show below that this linear model
(unlike the Chaplygin gas model) describes distinctively different
types of cosmological scenarios: the Big Bang, Big Crunch, Big
Rip, anti-Big Rip, solutions with de the Sitter attractor,
bouncing solutions, and their various combinations.

The paper is organized as follows. The principal part of the paper is
Sec.~\ref{Gen} in which the basic properties and particular
analytical solutions for a perfect fluid cosmology with a linear
equation of state are derived. We show that dark energy with the
equation of state $p=\alpha(\rho-\rho_0)$ may be effectively
reduced to the $\Lambda$-term and to the simplified linear
equation of state $p_{\rm eff}=w_{\rm eff}\rho_{\rm eff}$. In this
case, either the effective density $\rho_{\rm eff}$ or the density
of an effective $\Lambda$-term may in general have the negative
values. In dependence of signs of parameters $\alpha$ and $\rho_0$
the four different cosmological scenarios are considered. To study
the cosmological dynamics of the universe for different scenarios
the phase plane analysis is used. In Sec.~\ref{Matter} we discuss
the restrictions which can be set on this cosmological model from
the astronomical observations. The discussion of the results and
their further possible generalization is presented in
Sec.~\ref{Discussion}.

\section{Linear equation of state}

\label{Gen}

We consider the flat Friedman-Robertson-Walker universe filled
with a perfect fluid. For the sake of simplicity we will call
below this fluid as dark energy. Using the unit conventions $8\pi
G/3=c=1$ the corresponding Einstein equations for this cosmology
can be written as follows:
\begin{equation}
  \label{H^2}
  H^2=\rho_{\rm DE},
\end{equation}
\begin{equation}
  \label{rhoDE}
  \dot{\rho}_{\rm DE}=-3 H (\rho_{\rm DE}+p_{\rm DE}),
\end{equation}
where $H=\dot a/a$ is the Hubble parameter and $\rho_{\rm DE}$ and
$p_{\rm DE}$ are the energy density and pressure of a dark energy
correspondingly. In this paper we consider a perfect fluid with
the linear equation of state of the following general form:
\begin{equation}
  \label{pDE}
  p_{\rm DE}=\alpha(\rho_{\rm DE}-\rho_0),
\end{equation}
where $\alpha$ and $\rho_0$ are constants. Certain features of
cosmology with $\rho_0\ne 0$ and $0<\alpha<1$ was considered in
\cite{chiba}. The equation of state (\ref{pDE}) was used in
\cite{BDE} to describe the dark energy accretion onto the black
hole. Note the difference of this equation of
state (\ref{pDE}) from the commonly used one $p=w\rho$ with
$w=const$. Our simple generalization allows to include dark (and
also phantom) energy with a positive squared sound speed of linear
perturbations $c_s^2\equiv\partial p/\partial\rho=\alpha\geq0$.

\begin{figure}[t]
\begin{center}
\includegraphics[width=0.8\textwidth]{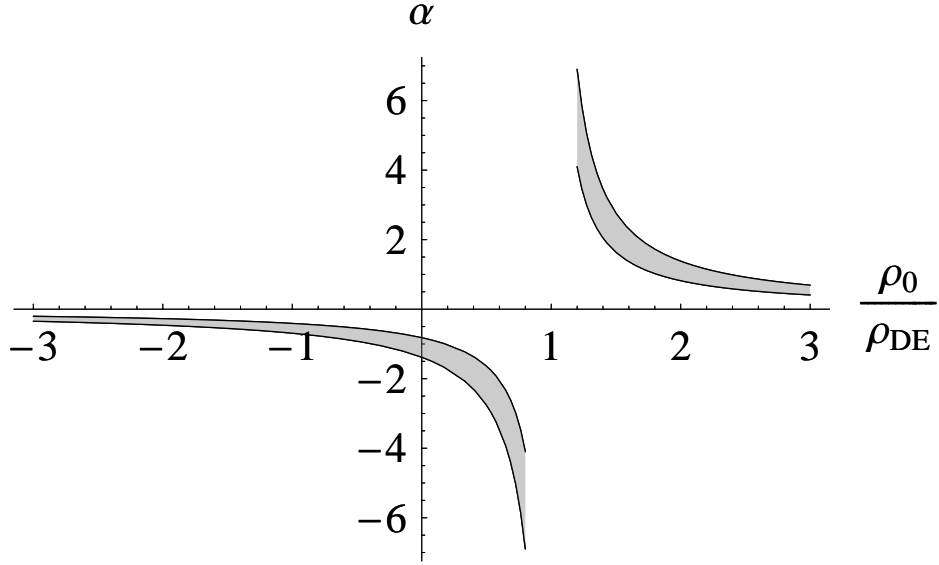}
\end{center}
\caption{\label{obs} Observational restrictions \cite{w} on the
allowable of the equation of state, $-1.38<w<-0.82$
recalculated and presented for parameters $\alpha$ and $\rho_0$
of the linear model of dark energy. Here $\alpha$ and $\rho_0$
are the parameters in (\ref{pDE}) and $\rho_{\rm DE}$ is the present
density of dark energy.}
\end{figure}

One can use the observational restrictions
\cite{w} on the allowable range of equation of state,
$-1.38<w<-0.82$, to put the restrictions on the parameters
of a linear model (\ref{pDE}). Substituting (\ref{pDE}) into
the equation of state $w=p_{\rm DE}/\rho_{\rm DE}$ we obtain
the range of allowable parameters parameters $\alpha$ and $\rho_0$,
shown in the Fig. (\ref{obs}).

The equation of state (\ref{pDE}) can be reduced to the
``effective'' cosmological constant and the
dynamically evolving dark energy
by redefining the fluid density and pressure in the following way:
\begin{eqnarray}
  \label{repl}
  \rho_{\rm DE}=\rho_\Lambda+\rho_\alpha, \quad
  p_{\rm DE}=p_\Lambda+p_\alpha,
\end{eqnarray}
where
\begin{equation}
  \label{rhoLN}
  p_\Lambda=-\rho_\Lambda, \quad p_\alpha=\alpha\rho_\alpha,
\end{equation}
and
\begin{equation}
  \label{rhoLN1}
  \rho_\Lambda=\frac{\alpha\rho_0}{1+\alpha}, \quad
  \rho_\alpha=\rho_{\rm DE}-\frac{\alpha\rho_0}{1+\alpha}.
\end{equation}
In Eqs. (\ref{repl},\ref{rhoLN},\ref{rhoLN1}) we denote
$\rho_\Lambda$ and $p_\Lambda$ as the density and the pressure of
the ''effective'' cosmological constant. Correspondingly, $\rho_\alpha$
and $p_\alpha$ are the density and pressure of the dynamically evolving
part of dark energy. The values $\rho_{\Lambda}$ and $p_{\Lambda}$
obviously obey the relation $\dot{\rho}_\Lambda=-3 H
(\rho_\Lambda+p_\Lambda)$ and therefore $\rho_\alpha$ and $p_\alpha$ obey
the equation:
\begin{equation}
  \label{rhom}
  \dot{\rho}_\alpha=-3 H (\rho_\alpha+p_\alpha).
\end{equation}
The cosmology with $\Lambda$-term and a linear equation of state
(for $p/\rho>-1$) was considered previously in \cite{Sil}. Note,
however, that in our consideration either $\rho_\Lambda$ or
$\rho_\alpha$ may be negative (nevertheless the sum
$\rho_\Lambda+\rho_\alpha$ must be positive) and also equation of state
for dynamically changing part of dark energy may be
super-negative. This leads to the new distinctive types of
cosmological scenarios considered below. The signs of $\rho_\alpha$ and
$\rho_\Lambda$ are conserved during the universe evolution as can
be seen from (\ref{rhom}) and (\ref{rhoDE}). Below we obtain the
full analytical solutions of evolution of the universe. In both
cases $\rho+p>0$ and $\rho+p<0$ we find from (\ref{rhom}) and
(\ref{rhoDE}):
\begin{equation}
  \label{rho}
  \rho_\alpha=\frac{A}{a^{3(1+\alpha)}},
\end{equation}
where constant $A$ may be either positive or negative. Two different
asymptotic regimes are possible: for $|\rho_\alpha|\ll|\rho_\Lambda|$
the universe behaves like the de Sitter universe with
\begin{equation}
  \label{asympt1}
  \rho \approx \rho_\Lambda, \quad p\approx -\rho_\Lambda,
\end{equation}
and for $|\rho_\alpha|\gg|\rho_\Lambda|$ the universe is filled with
dynamical part of dark energy:
\begin{equation}
  \label{asympt2}
  \rho\approx\rho_\alpha.
\end{equation}
Using (\ref{H^2}) and (\ref{rho}) we obtain the relation between
the differentials $da$ and $dt$:
\begin{equation}
  \label{difat}
  \frac{da}{a\sqrt{\rho_\Lambda+A a^{-3(1+\alpha)}}}
  =\pm dt.
\end{equation}
In dependence on signs of $\rho_\Lambda$ and $A$ there can be
three different results for integration of relation (\ref{difat}).
For $\rho_\Lambda>0$ and $A>0$ we find:
\begin{equation}
  \label{a1+}
  a(t)=
  \left\{\sqrt{\beta}\;
  \sinh\left[\pm\frac{\kappa}{2}\sqrt{\rho_\Lambda}\;t\right]
  \right\}^{2/\kappa}.
\end{equation}
Here we denote $\kappa=3(1+\alpha)$ and $\beta=A/\rho_\Lambda$.
The choice of the upper or lower sign in (\ref{a1+}) depends on
the sign ''$+$'' or ''$-$'' in (\ref{difat}). The expression
(\ref{a1+}) was first obtained in \cite{Sil}. The asymptotic
behavior of (\ref{a1+}) for $t\to 0$ is given by
\begin{equation}
  \label{a1+0}
  a=\left(\pm\frac{\kappa}{2}\sqrt{A}\;t\right)^{2/\kappa},
\end{equation}
The Eq. (\ref{a1+}) for $t\to\pm\infty$ can be rewritten as
follows:
\begin{equation}
  \label{a1+infty}
  a=\left(\frac{|\beta|}{4}\right)^{1/\kappa}
  \exp\left(\pm\sqrt{\rho_\Lambda}\;t\right),
\end{equation}
The expression (\ref{a1+infty}) is valid only if $\kappa>0$. For
$\kappa<0$ the signs in exponent in this expression should be
changed to opposite ones.

In the case $\rho_\Lambda>0$ and $A<0$ the integration of
(\ref{difat}) gives:
\begin{equation}
  \label{a2}
  a(t)=
  \left\{\sqrt{-\beta}\;
  \cosh\left[\frac{\kappa}{2}\sqrt{\rho_\Lambda}\;t\right]
  \right\}^{2/\kappa}.
\end{equation}
In slightly different form the solution (\ref{a2}) was obtained in
\cite{McInnes01} for the restricted case $-1<\alpha<0$. The
asymptotic behavior of a scale factor for $t\to 0$ is given by:
\begin{equation}
  \label{a20}
  a(t)=(-\beta)^{1/\kappa}
  \left(1+\frac{\kappa}{4}\;\rho_\Lambda\;t^2\right).
\end{equation}
For $t\to\pm\infty$ the scale factor evolution is described by
(\ref{a1+infty}). Finally, for $\rho_\Lambda<0$ and $A>0$ one can
find the evolution of the scale factor of the universe:
\begin{equation}
  \label{a3+}
  a(t)=
  \left\{\sqrt{-\beta}\;
  \sin\left[\frac{\kappa}{2}\sqrt{-\rho_\Lambda}\;t\right]
  \right\}^{2/\kappa}.
\end{equation}
In the limitation $-1<\alpha<1$ the similar solution was found in
\cite{McInnes04} for the cosmology with negative anti-de Sitter
$\Lambda$-term plus a scalar quintessential field with special
form of a potential. For $t\to 0$ the asymptotic behavior of scale
factor is given by (\ref{a1+0}). The behavior of a scale factor
near $t\simeq \pi/(\kappa\rho_\Lambda)$ can be described as
follows:
\begin{equation}
  \label{a3+pi.2}
  a(t)=\left\{\sqrt{-\beta}\;
 \left[1-\frac{1}{2}\left(\frac{\kappa}{2}
 \sqrt{-\rho_\Lambda}\;t-\frac{\pi}{2}\right)^2\right]
  \right\}^{2/\kappa}.
\end{equation}
Correspondingly, the asymptotic behavior of scale factor at
$t\to\pm 2\pi/(\kappa\rho_\Lambda)$ is given by
\begin{equation}
  \label{a3+pi}
  a=
  \left(\pi\sqrt{-\beta}-\frac{\kappa}{2}\sqrt{A}\;t\right)^{2/\kappa}.
\end{equation}
To find the attractors and unstable points of the solution of the
Friedman equations we use the method of the phase trajectories.
Denoting $y=\rho_\alpha+p_\alpha$ we obtain the following system of
equations
\begin{equation}
  \label{PT1}
  \dot H=-3y/2, \quad  \dot y=-3(\alpha+1)Hy.
\end{equation}
All phase trajectories of the system lie on the curve
$H^2=\rho_{\rm DE}$ which can be rewritten as
$y=(\alpha+1)H^2-\alpha\rho_0$. The particular form of this curve
and direction of the phase trajectory depends on the signs of
$1+\alpha$ and $\alpha\rho_0$. The four cases are possible:

(i) $1+\alpha>0$ and $\alpha\rho_0>0$.

\begin{figure}[t]
\begin{tabular}{c c}
\includegraphics[angle=0,width=0.45\textwidth]{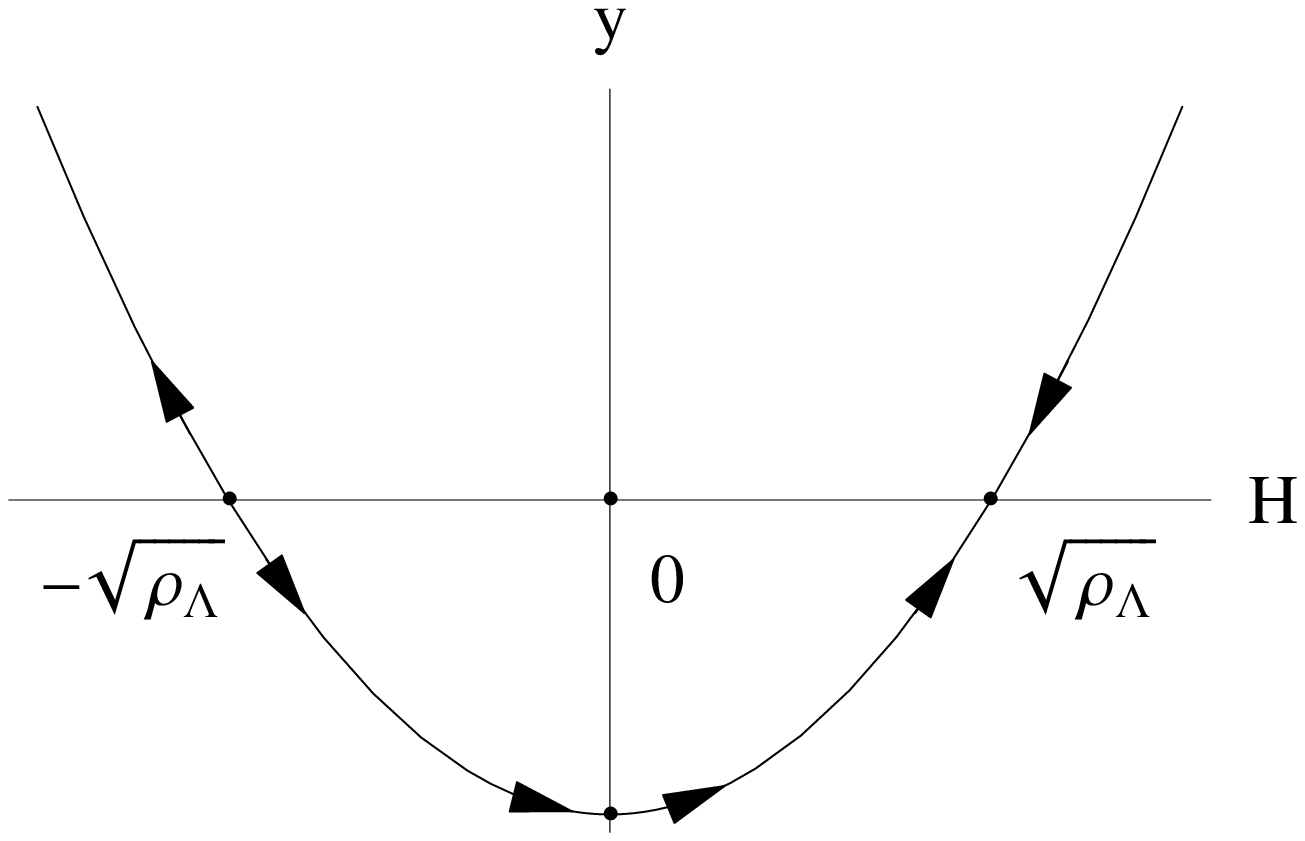}&
\includegraphics[angle=0,width=0.45\textwidth]{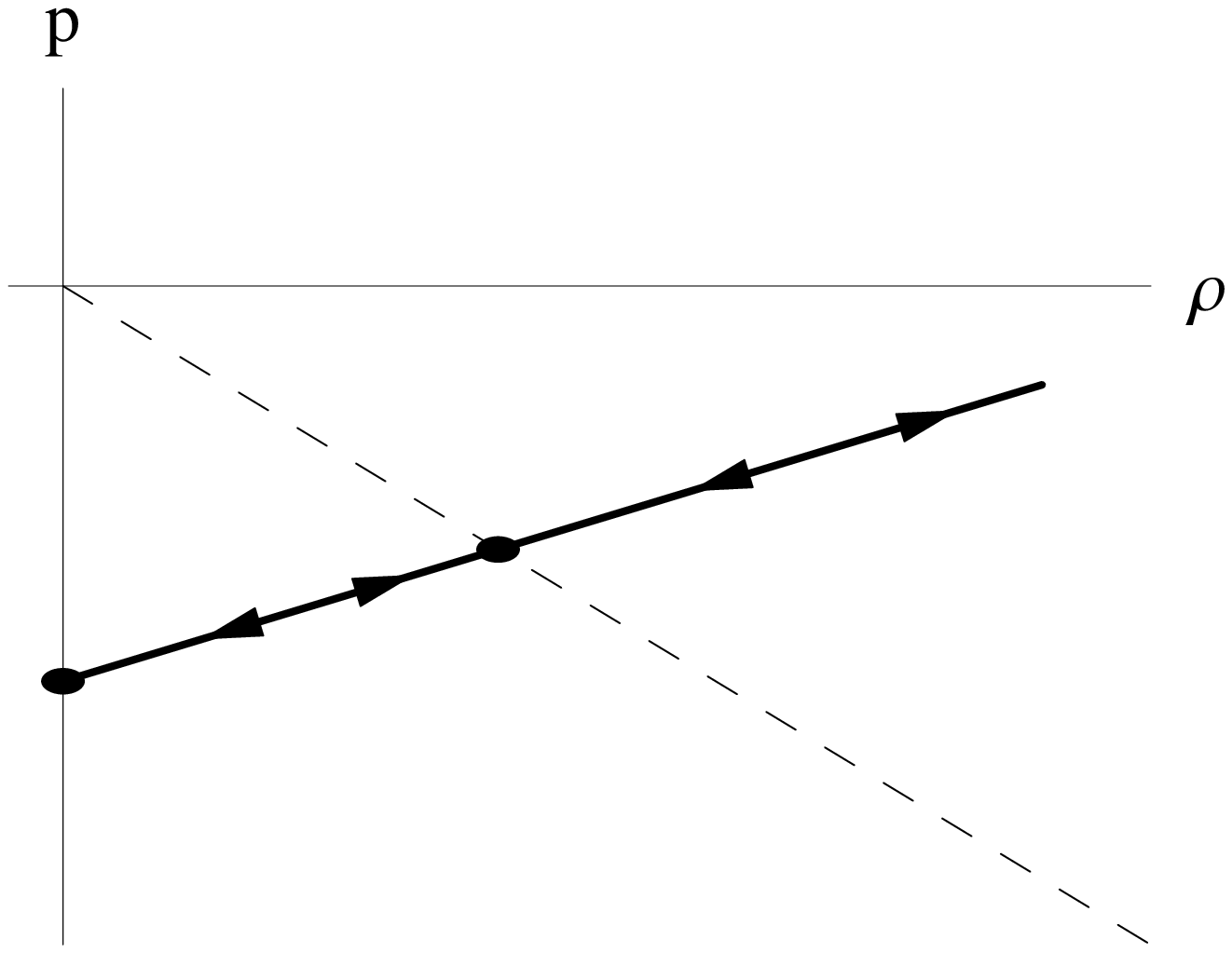}\\
\end{tabular}
\caption{
 \label{case1}
Phase trajectory of the universe and diagram of evolution in
coordinates ($p$, $\rho$) are shown in the case of dark energy
parameters $1+\alpha>0$ and $\alpha\rho_0>0$. In the left panel
the phase space diagram of evolution in coordinates $H$ and
$y\equiv \rho_\alpha+p_\alpha$ is presented. The direction of evolution
is shown by arrows. There are two unstable
singular points, ($H= -\rho_{\Lambda}^{1/2}$, $y=0$) and
($H=0$, $y= -\alpha\rho_0$), and one attractor
($H=\rho_{\Lambda}^{1/2}$, $y=0$). In the right panel the
evolution of is shown in coordinates $\rho$ and $p$.
The evolution from $\rho=\infty$ to $\rho=\rho_\Lambda$
corresponds to the right branch
of parabola ($y>0$, $H>0$) of the phase space diagram. The reverse
process (from $\rho=\rho_\Lambda$ to $\rho=\infty$) corresponds
to the left branch of parabola ($y>0$, $H<0$) of phase space diagram. 
The evolution from $\rho=\rho_\Lambda$ to bounce $\rho=0$ and
back to $\rho=\rho_\Lambda$ corresponds to the middle branch
of parabola ($y<0$) of phase space diagram.}
\end{figure}

There are two singular points at $y=0$ axis. At the point
($H=\rho_{\Lambda}^{1/2}$, $y=0$) the linearized system of
evolution equations is
\begin{equation}
\dot H=-3y/2, \quad \dot y=-3(1+\alpha)\rho_{\Lambda}^{1/2}y,
\end{equation}
The first eigenvalue and the first eigenvector of the system equal to zero. 
The second
eigenvalue equals to $\lambda_2=3(1+\alpha)\rho_{\Lambda}^{1/2}<0$ and the
second eigenvector is $\{3/2, 3(1+\alpha)\rho_{\Lambda}^{1/2}\}$.  
We can see that the considered point is the attractor.
The same linearization method is used for the universe evolution
analysis near other singular points. The second singular
point ($H= -\rho_{\Lambda}^{1/2}$, $y=0$) corresponds to the
unstable equilibrium state. The another interesting singular point
is ($H=0$, $y= -\alpha\rho_0$), in which the universe reaches the
zero density $\rho_{\rm DE}=0$ and its contraction is changed to
the expansion. The phase trajectories near the singular points are
plotted in the left panel of Fig.~\ref{case1} with the directions of
evolution marked by arrows. In the auxiliary right panel of
Fig.~\ref{case1} the evolution of pressure $p$ as a function of
the dark energy density $\rho$ is shown.

The right branch of parabola in the left panel of Fig.~\ref{case1}
($y>0$, $H>0$) corresponds to the solution (\ref{a1+}) with an upper
sign at $t>0$. In this case the universe is filled with
non-phantom energy. The universe starts from the Big Bang,
corresponding to the initial values of density $\rho_{{\rm
DE},i}=+\infty$ and scale factor $a_{i}=0$. The asymptotic
behavior of a scale factor at $a\to 0$ is described by
(\ref{a1+0}) with an upper sign. The relation (\ref{a1+0}) with an upper
sign describes also the evolution of a scale factor for
$\rho_\Lambda=0$ for all $t$. At the late times the universe
approaches to the de Sitter regime (\ref{asympt1}). The sign of
$\rho+p$ is conserved during the evolution of the universe. The
asymptotic behavior of $a$ for $t\to\infty$ is given by
(\ref{a1+infty}) with an  upper sign.
The left branch of parabola in left panel of Fig.~\ref{case1}
($y>0$, $H<0$) corresponds to the reverse process with respect to
the examined above and is described by (\ref{a1+}) with a lower sign.
Note that $t$ is taken to be negative in this case. The asymptotic
behavior for $t\to 0$ and ($t\to\infty$) is given by (\ref{a1+0})
and (\ref{a1+infty}) correspondingly with lower sighs in both
cases.
A middle part of the parabola in left panel of Fig.~\ref{case1} ($y<0$)
corresponds to the solution (\ref{a2}). This particular case was
considered in \cite{McInnes01} as a simplest example of phantom
cosmology without a Big Rip. The universe in this case is filled
with a phantom energy. Because of the specifics of a linear
equation of state, the considered universe with a phantom energy
is not born in the Big Bang. Instead of this, the universe starts
in this case from the initial scale factor $a_{i}=+\infty$. For
$t\to -\infty$ it behaves like the de Sitter universe (however
reversed in time) according to (\ref{a1+infty}) with an upper sign and bounces
at the minimal value of a scale factor $a_{min}=
(-\beta)^{1/\kappa}$. In this moment the state of dark energy is
very special: the pressure is finite but the total energy density
is zero, $\rho_\alpha+\rho_\Lambda=0$. Near the bounce this universe can
be described by (\ref{a20}). After the bouncing the universe
expands and at the late times it approaches to the de Sitter state
(\ref{a1+infty}) with an upper sign.

(ii) $1+\alpha>0$ and $\alpha\rho_0<0$.

\begin{figure}[t]
\begin{tabular}{c c}
\includegraphics[angle=0,width=0.45\textwidth]{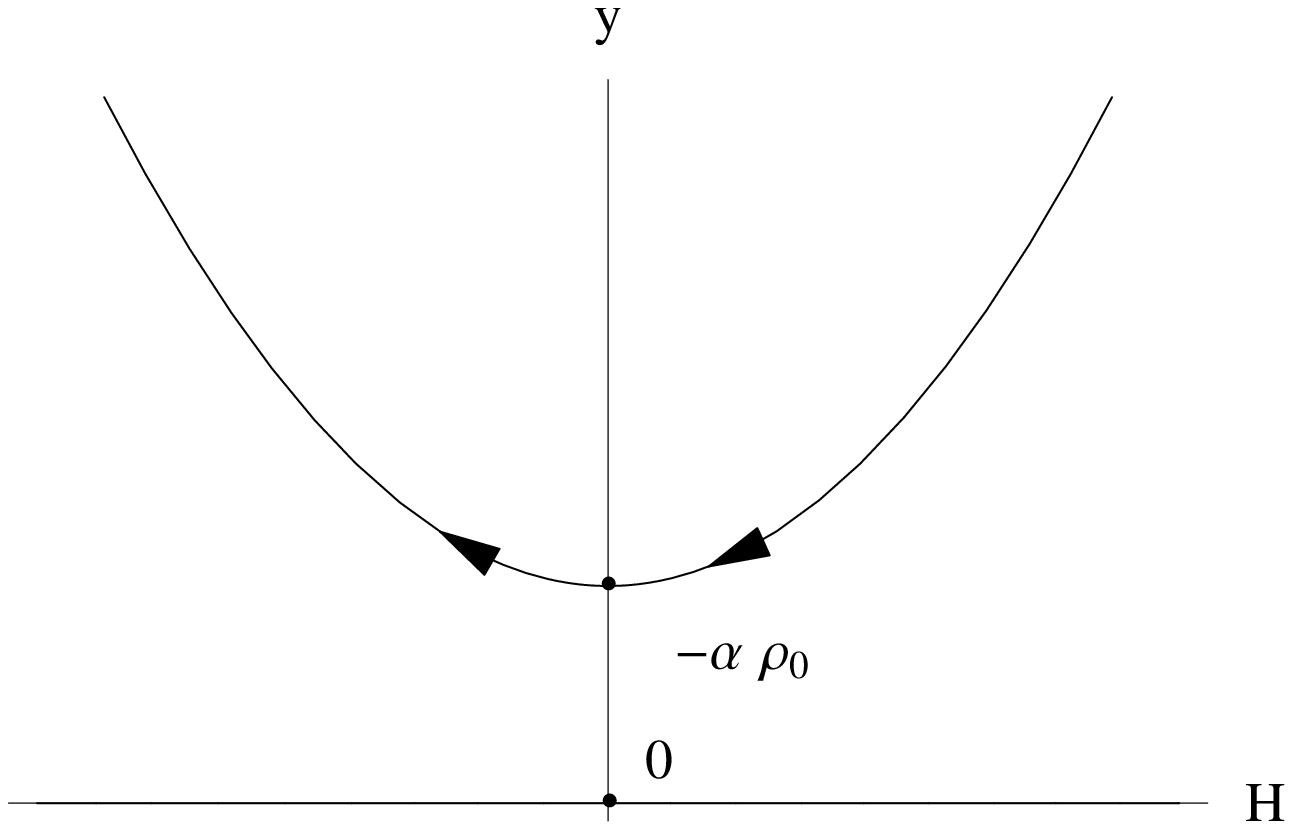}&
\includegraphics[angle=0,width=0.45\textwidth]{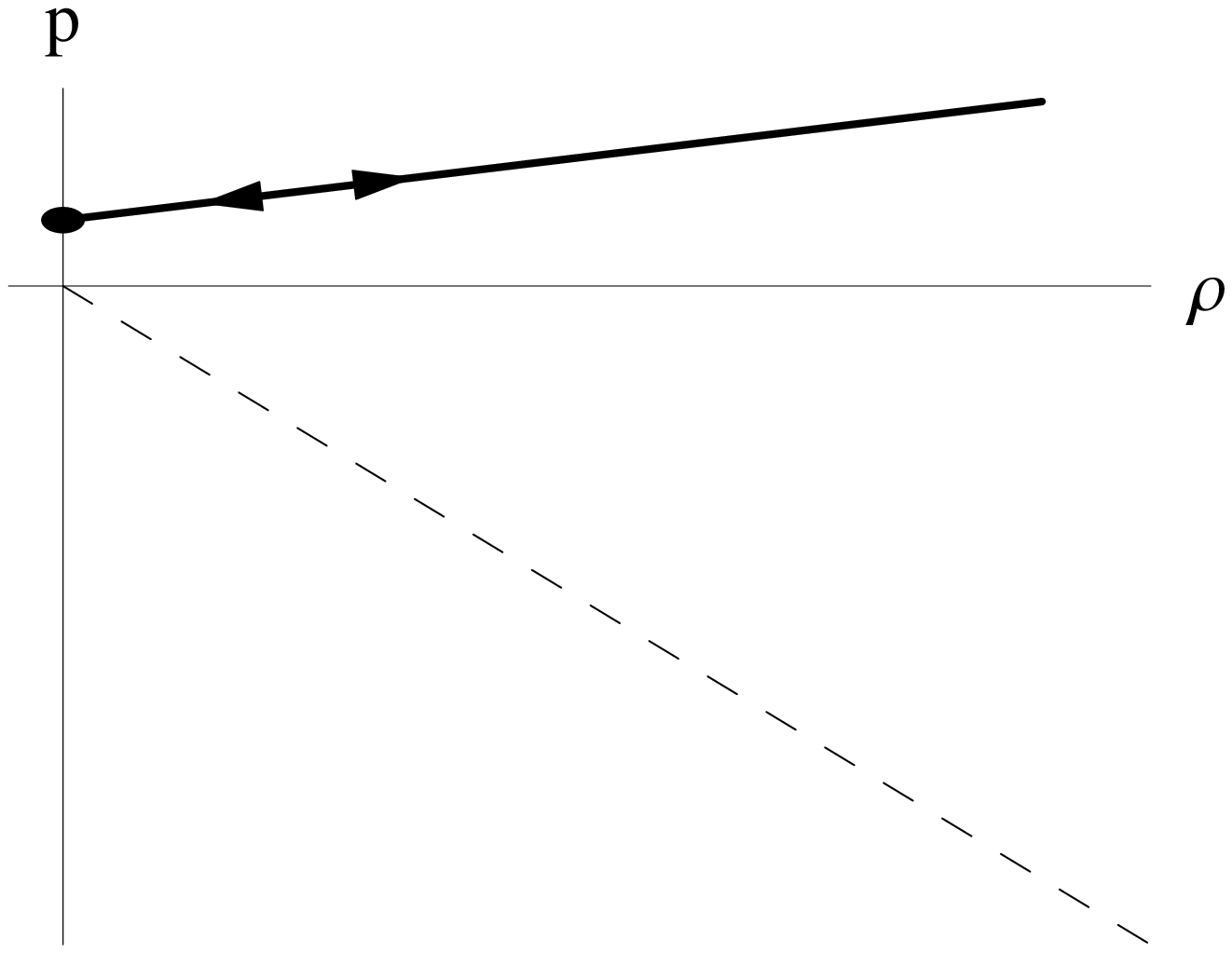}\\
\end{tabular}
\caption{
 \label{case2}
Phase trajectory of the universe and diagram of evolution in
coordinates ($p$, $\rho$) are shown in the case of dark energy
parameters $1+\alpha>0$ and  $\alpha\rho_0<0$. In the left panel
the phase space diagram of evolution in coordinates $H$ and
$y\equiv \rho_\alpha+p_\alpha$ is presented. The direction of evolution
is shown by arrows. There is only one singular point,
($H=0$, $y= -\alpha\rho_0$), which is unstable. In the right panel the
evolution is shown in coordinates $\rho$ and $p$.
The universe evolves from $\rho=\infty$ to $\rho=0$, which corresponds
to the right branch
of parabola ($H>0$) of the phase space diagram. At the point $\rho=0$
(corresponding to the point ($H=0$, $y= -\alpha\rho_0$) of the phase
space diagram) there is a bounce. Then the universe evolves back to
$\rho=\infty$ which corresponds
to the left branch of parabola ($H<0$) of the phase space diagram.}
\end{figure}

The density of the $\Lambda$-term is negative, $\rho_{\Lambda}<0$.
The linearized system of the evolution equations at the point
($H=0$, $y=-\alpha\rho_0$) is 
\begin{equation}  \dot
H=3\alpha\rho_0/2-3(y+\alpha\rho_0)/2, \quad \dot
y=3(1+\alpha)\alpha\rho_0H. 
\end{equation} 
Universe expands starting from the Big Bang, reaches the zero density at the point
($H=0$, $y=-\alpha\rho_0$) and then its expansion changes to the
contraction. The phase trajectory and diagram of evolution
$p-\rho$ are shown in the Fig.~\ref{case2} (the similar
cosmological behavior was obtained in \cite{McInnes04} for the
universe with negative $\Lambda$-term and quintessence with
special potential). The universe is filled with non-phantom energy
and starts from the Big Bang. The evolution of a scale factor is
described by (\ref{a3+}) and at early times by the asymptotic
(\ref{a1+0}) with an upper sign. This asymptotic is also valid
for $\rho_\Lambda=0$ at all $t$. During the finite time $\Delta
t=\pi/(\kappa\rho_\Lambda)$ a scale factor $a$ reaches the
maximum value $a_{max}=(-\beta)^{1/\kappa}$ at which the universe bounces.
Near the bounce the behavior of a scale factor can be described
by (\ref{a3+pi.2}). After the bounce the
universe contracts and in a finite time $\Delta t$ it collapses to
the Big Crunch. The scale factor of the universe near the collapse
is given by (\ref{a3+pi}).

(iii) $1+\alpha<0$ and $\alpha\rho_0<0$.

\begin{figure}[t]
\begin{tabular}{c c}
\includegraphics[angle=0,width=0.45\textwidth]{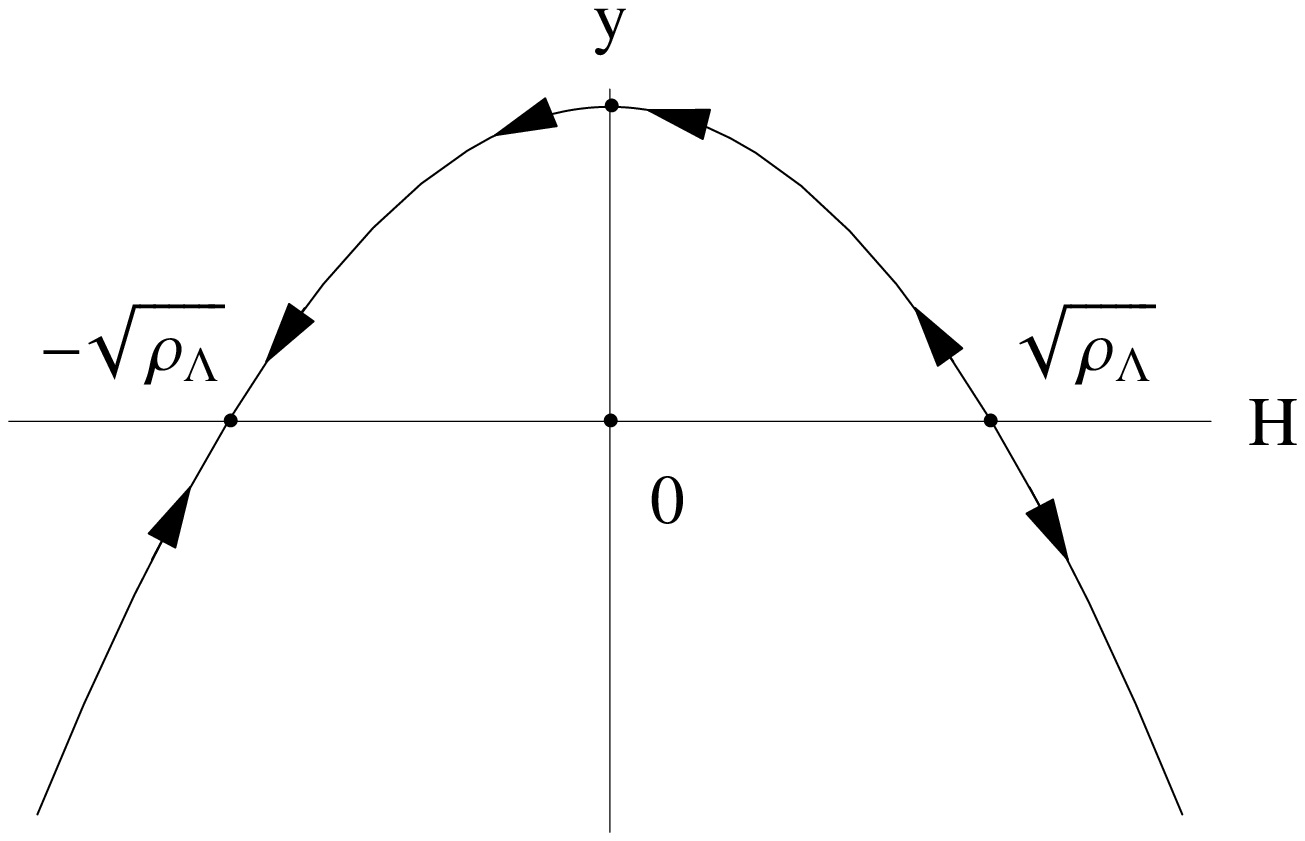}&
\includegraphics[angle=0,width=0.45\textwidth]{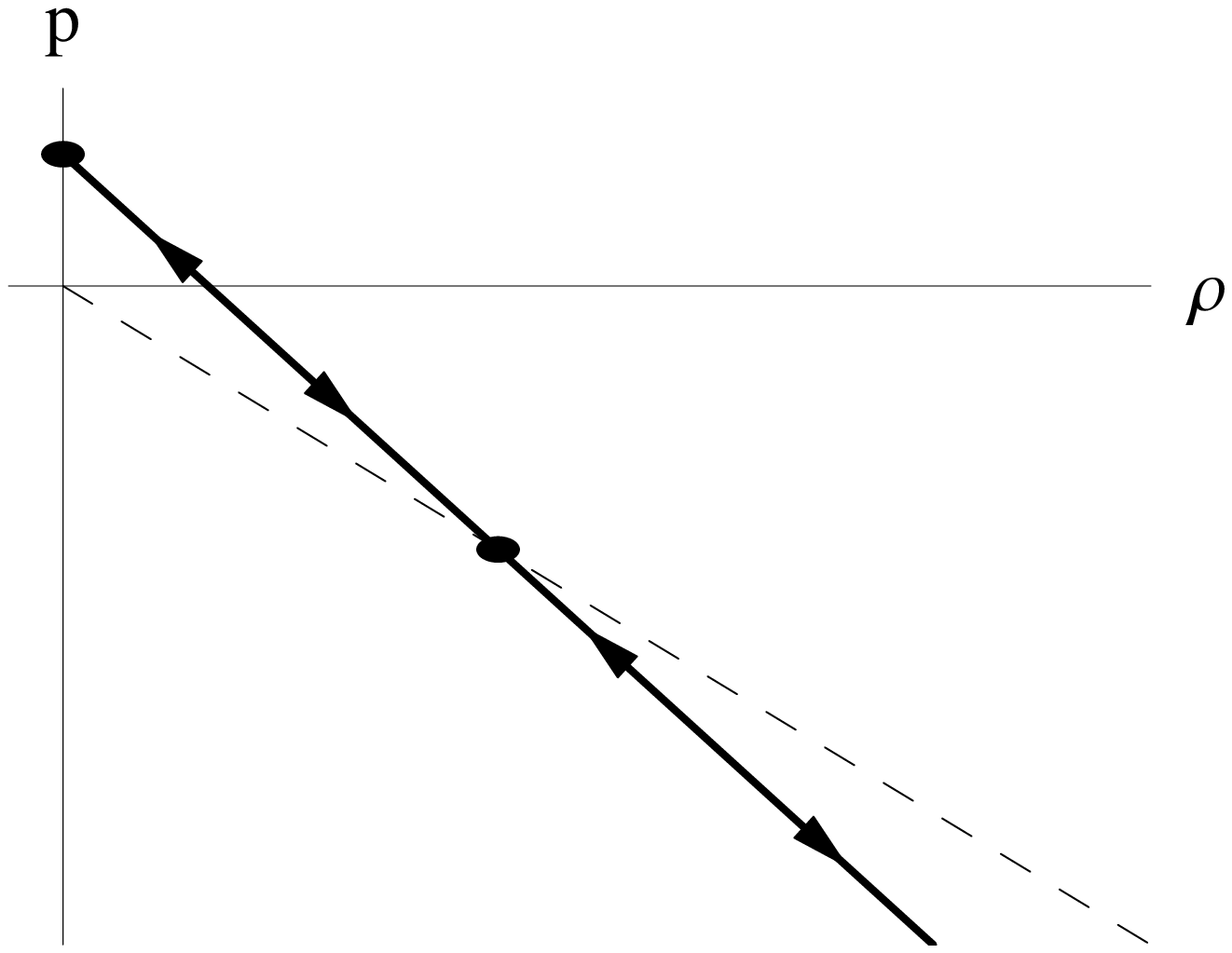}\\
\end{tabular}
\caption{
 \label{case3}
Phase trajectory of the universe and diagram of evolution in
coordinates ($p$, $\rho$) are shown in the case of dark energy
parameters $1+\alpha<0$ and $\alpha\rho_0<0$. In the left panel
the phase space diagram of evolution in coordinates $H$ and
$y\equiv \rho_\alpha+p_\alpha$ is presented. The direction of evolution
is shown by arrows. There are two unstable
singular points, ($H= \rho_{\Lambda}^{1/2}$, $y=0$) and
($H=0$, $y= -\alpha\rho_0$), and one attractor
($H=-\rho_{\Lambda}^{1/2}$, $y=0$). In the right panel
evolution of the universe is shown in coordinates $\rho$ and $p$.
The evolution from $\rho=\rho_\Lambda$ to $\rho=\rho_\infty$
corresponds to the right branch
of parabola ($y<0$, $H>0$) of the phase space diagram. The reverse
process from $\rho=\infty$ to $\rho=\rho_\Lambda$ corresponds
to the left branch of parabola ($y<0$, $H<0$) of phase space diagram. 
The evolution from the $\rho=\rho_\Lambda$ to bounce at $\rho=0$ and
back to $\rho=\rho_\Lambda$ corresponds to the middle branch
of parabola ($y>0$) of phase space diagram.}
\end{figure}

At the zero density in the point ($H=0$, $y=-\alpha\rho_0$) the
universe expansion is changed to the contraction. The singular
point ($H=-\rho_{\Lambda}^{1/2}$, $y=0$) is an attractor and
respectively ($H=\rho_{\Lambda}^{1/2}$, $y=0$) is an unstable
equilibrium point. The phase trajectory and diagram of evolution
$p-\rho$ are shown in the Fig.~\ref{case3}.

First we consider the right branch of the parabola ($y<0$, $H>0$).
In this case the universe is filled with phantom energy,
$\rho+p<0$. The evolution of a scale factor is given by
(\ref{a1+}) with an upper sign. Note that $t$ is taken to be
negative in this case. Starting from the scale factor $a_{i}=0$ the
universe expands during an infinite time to the final Big Rip. At
$t\to-\infty$ a scale factor of the universe is given by
(\ref{a1+infty}) with an upper sign. Near the Big Rip a scale
factor $a$ is described by (\ref{a1+0}) with an upper sign
(note that $\kappa$ and $t$ are both negative).
The left branch of the parabola ($y<0$, $H<0$) corresponds to the
reverse process with respect to the examined above and is
described by (\ref{a1+}) with a lower sign. Time $t$ is positive
during the whole evolution. In this case the universe is filled
with phantom energy and it starts from the ''anti-Big Rip''
solution (with the infinite value of the initial scale factor),
contracts during an infinite time to the final $a_f=0$.
The asymptotic behavior for $t\to 0$ and $t\to+\infty$ is given by
(\ref{a1+0}) with the lower sign and (\ref{a1+infty}) with the
lower sign correspondingly.
The middle part of the parabola ($y>0$) corresponds to the
solution (\ref{a2}). The universe is filled with a non-phantom
energy and starts from $a_{i}=0$. At $t\to -\infty$
the scale factor of the universe is given by (\ref{a1+infty}) with
the upper sign. After the bouncing at the maximum value of a scale
factor $a_{max}= (-\beta)^{1/\kappa}$ the universe begins to
contract. Near the bouncing point the scale factor is given by
(\ref{a20}). At $t\to +\infty$ the universe behaves like
(\ref{a1+infty}) with the lower sign.

(iv) $1+\alpha<0$ and $\alpha\rho_0>0$.

\begin{figure}[t]
\begin{tabular}{c c}
\includegraphics[angle=0,width=0.45\textwidth]{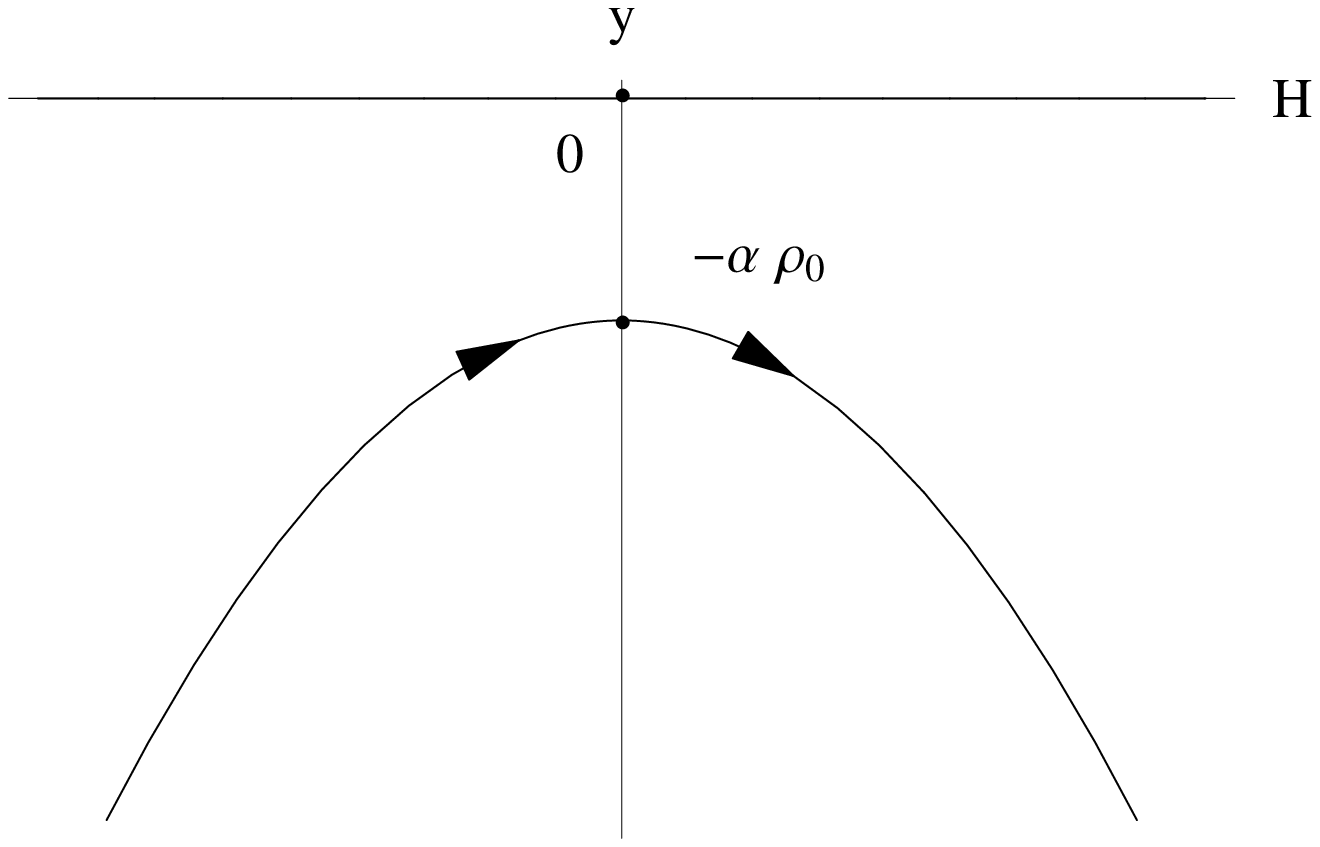}&
\includegraphics[angle=0,width=0.45\textwidth]{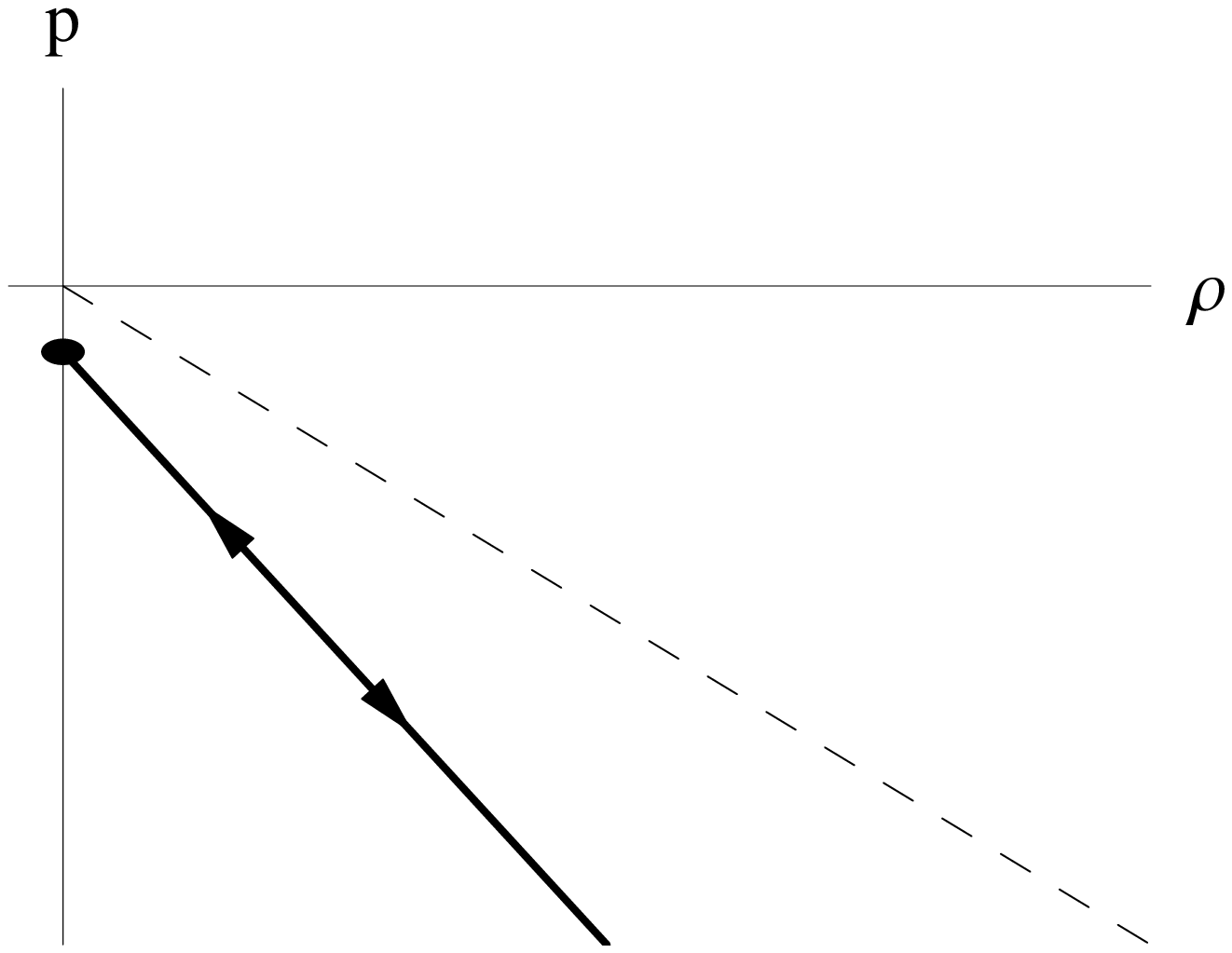}\\
\end{tabular}
\caption{
 \label{case4}
Phase trajectory of the universe and diagram of evolution in
coordinates ($p$, $\rho$) are shown in the case of dark energy
parameters $1+\alpha<0$ and   $\alpha\rho_0>0$. In the left
panel the phase space diagram of evolution in
coordinates $H$ and $y\equiv \rho_\alpha+p_\alpha$ is presented.
The direction of evolution is shown by arrows. There is only one
singular point, ($H=0$, $y= -\alpha\rho_0$), which is unstable. In
the right panel the evolution of the universe is shown in
coordinates $\rho$ and $p$. The universe evolves from ''anti-Big
Rip'' state ($\rho=\infty$) to the bouncing point ($\rho=0$),
which corresponds to the left branch of parabola ($H<0$) of the
phase space diagram. Then the universe evolves to the Big Rip ($\rho=\infty$) which
corresponds to the right branch of parabola ($H>0$) of the phase
space diagram.}
\end{figure}

In this case the density of the $\Lambda$-term is negative:
$\rho_{\Lambda}<0$. The universe reaches the zero density at the
point ($H=0$, $y=-\alpha\rho_0$), where the contraction is changed
to the expansion. The phase trajectory and diagram of evolution
$p-\rho$ is shown in the Fig.~\ref{case4}. The universe is filled
with a phantom energy and the solution for a scale factor is given
by (\ref{a3+}). The universe is born in the ''anti-Big Rip'' state
and the scale factor is given by (\ref{a1+0}) with the lower sign
near $t\approx 0$. Then the universe contracts to the minimum
state $a_{min}=(-\beta)^{1/\kappa}$ in a finite time $\Delta
t=\pi/(\kappa\rho_\Lambda)$, and after bouncing it begins to
expand. In time $\Delta t$ the universe comes to the Big Rip where
the scale factor of the universe is given by (\ref{a3+pi}).

Different cosmological scenarios discussed above in this Section
are summarized in Table~\ref{tab1}.

\section{Universe with dark energy and matter}

\label{Matter}

If we add the usual matter (dark matter, baryons and radiation) to
the considered universe then the differential equation for a scale
factor evolution would have the following form:
\begin{equation}
  \label{difat1}
\left(\frac{\dot a}{a}\right)^2=H_0^2
[\Omega_{r,0}(1+z)^4+\Omega_{m,0}(1+z)^3 + \Omega_{\Lambda,0} +
\Omega_{\alpha,0}(1+z)^{3(\alpha+1)}],
\end{equation}
where $H_0$ is the current value of the Hubble constant,
$z=a_0/a-1$ is a redshift, $\Omega_{r,0}$ and
$\Omega_{m,0}\simeq0.3$ are the cosmological density parameters of
radiation and non-relativistic matter correspondingly. The value
of the cosmological density parameter of ''$\Lambda$-term'' is:
$\Omega_{\Lambda,0}=[\alpha/(\alpha+1)]\rho_0/\rho_{c,0}$, where
$\rho_{c,0}$ is the current critical density. The contribution of
the dynamically changing part of dark energy is the following:
$\Omega_{\alpha,0} = \Omega_{q,0}-\Omega_{\Lambda,0}$, where
$\Omega_{q,0}\simeq0.7$ is the quintessence density parameter. We
suppose that the universe is flat with $\Omega_{r,0}+
\Omega_{m,0}+\Omega_{q,0}=1$. In the case of pure $\Lambda$-term
($\alpha=-1$, $\rho_0=0$), one should omit the last term in the
brackets in (\ref{difat1}) and take
$\Omega_{\Lambda,0}=\Omega_{q,0}$.

\begin{table}[t]
\caption{\label{tab1} Possible cosmological scenarios
in the case of a generalized linear equation of state 
$p=\alpha(\rho-\rho_0)$ depending on parameters 
$\alpha$ and $\rho_0$. \\ } 
\lineup {
\small
\begin{center}
\begin{tabular}{|p{3 cm}|p{2.3cm}|p{2.3cm}|p{2.3cm}|p{2.3cm}|}
 \hline Values of \newline $H=\dot a/a$ & 
   $1+\alpha>0$,\newline $\alpha\rho_0>0$ & 
   $1+\alpha>0$, \newline $\alpha\rho_0<0$ & 
   $1+\alpha<0$,\newline $\alpha\rho_0<0$ & 
   $1+\alpha<0$, \newline $\alpha\rho_0>0$\\ 
\hline $H<-\sqrt{\rho_{\Lambda}}$ & Contraction &  & Contraction\newline &\\
\cline{1-2}\cline{4-4} $H=-\sqrt{\rho_{\Lambda}}$ & 
  Non-steady\newline equilibrium point & 
  Expansion,\newline bounce and \newline contraction   
  & Attractor & Contraction \newline from the\newline anti-Big Rip,\\
\cline{1-2}\cline{4-4} $-\!\sqrt{\rho_{\Lambda}}\!<\!H\!<\!\sqrt{\rho_{\Lambda}}$&
  Contraction,\newline bounce and\newline expansion&  & 
  Expansion,\newline bounce and \newline contraction  
  &bounce and \newline expansion to\newline the Big Rip\\
\cline{1-2}\cline{4-4} $H=\sqrt{\rho_{\Lambda}}$& 
  Attractor& & Non-steady\newline equilibrium\newline point&\\
\cline{1-2}\cline{4-4} $H>\sqrt{\rho_{\Lambda}}$& Expansion \newline & & Expansion &\\
\hline
\end{tabular}
\end{center}
}
\end{table}

The observational restrictions on the $w$ ($-1.38<w<-0.82$),
derived from the SN data, limit only the behavior of the universe
at small redshifts $z\simeq0-1$. In the case of $\rho_0/\rho_{\rm DE}<1$
(see Fig.~\ref{obs}) the dark energy does not change
drastically the evolution of the universe at high redshift 
in comparison with a pure $\Lambda$-term case. 
The only restriction is the age of the
universe which is limited (e.g. by the oldest globular clusters) to
$t_0>12\cdot10^9$~yr. We find the age of the universe in our model by
integration of (\ref{difat1}) from $z=0$ to $z=\infty$ for
different values of $\alpha$ and $\rho_0$ and for nowadays values
of $H_0=65$~km~s$^{-1}$~Mpc$^{-1}$, $\Omega_m=0.3$ and 
$\Omega_r=4.54\cdot10^{-5}/h^2$, where $h$ is the $H_0$ in the units
$100$~km~s$^{-1}$~Mpc$^{-1}$. The connection of commonly used
parameter $w$ with our parameters $\alpha$ and $\rho_0$ is given
by relation $w=\alpha(1-\rho_0/\rho_{\rm DE})$. For example, if
$w=-0.82$ then the age restriction is important only at 
$\rho_0/\rho_{\rm DE}\leq-100$. In the case of $w<-1$ 
the universe is older then in a pure $\Lambda$-term case ($w=-1$) 
and the age restriction is unimportant. 

In contrast, at $\rho_0/\rho_{\rm DE}>1$ only a small
range of parameters may correspond to the real universe. In
addition to restriction on $w$ and the universe age restriction one must require
the universe to start from the Big Bang and not from the bounce
at a finite scale factor. The bounce $\dot a=0$ appears in the
case of $\alpha>1/3$ and $\Omega_{\alpha,0}<0$. Therefore, to
avoid the bounce we must additionally require that
$\alpha\leq1/3$. One more condition $\Omega_{\alpha,0}+ \Omega_{r,0}>0$
should be satisfied in the boundary case of $\alpha=1/3$. 
Strictly speaking, the bounce at redshift
$1+z_b=(\Omega_{r,0}/|\Omega_{\alpha,0}|)^{1/(3\alpha-1)}$ could not 
occur later then the end of inflation or reheating
moment at the temperature $T_{\rm RH}\sim10^{13}$~GeV 
corresponding to $z_{\rm RH}\sim 10^{26}$. Therefore it is
necessary to satisfy the condition $z_b\gg z_{\rm RH}$. 
The last condition puts the
restriction to the allowable values of $\alpha$ and
$\Omega_{\alpha,0}$. This restriction can be
considered in the particular inflation model with taking into
account the additional dynamical components (say inflaton). 
This consideration is out of scope of this paper.

\section{Discussion and Conclusion}

\begin{figure}[t]
\begin{tabular}{c c}
\includegraphics[angle=0,width=0.45\textwidth]{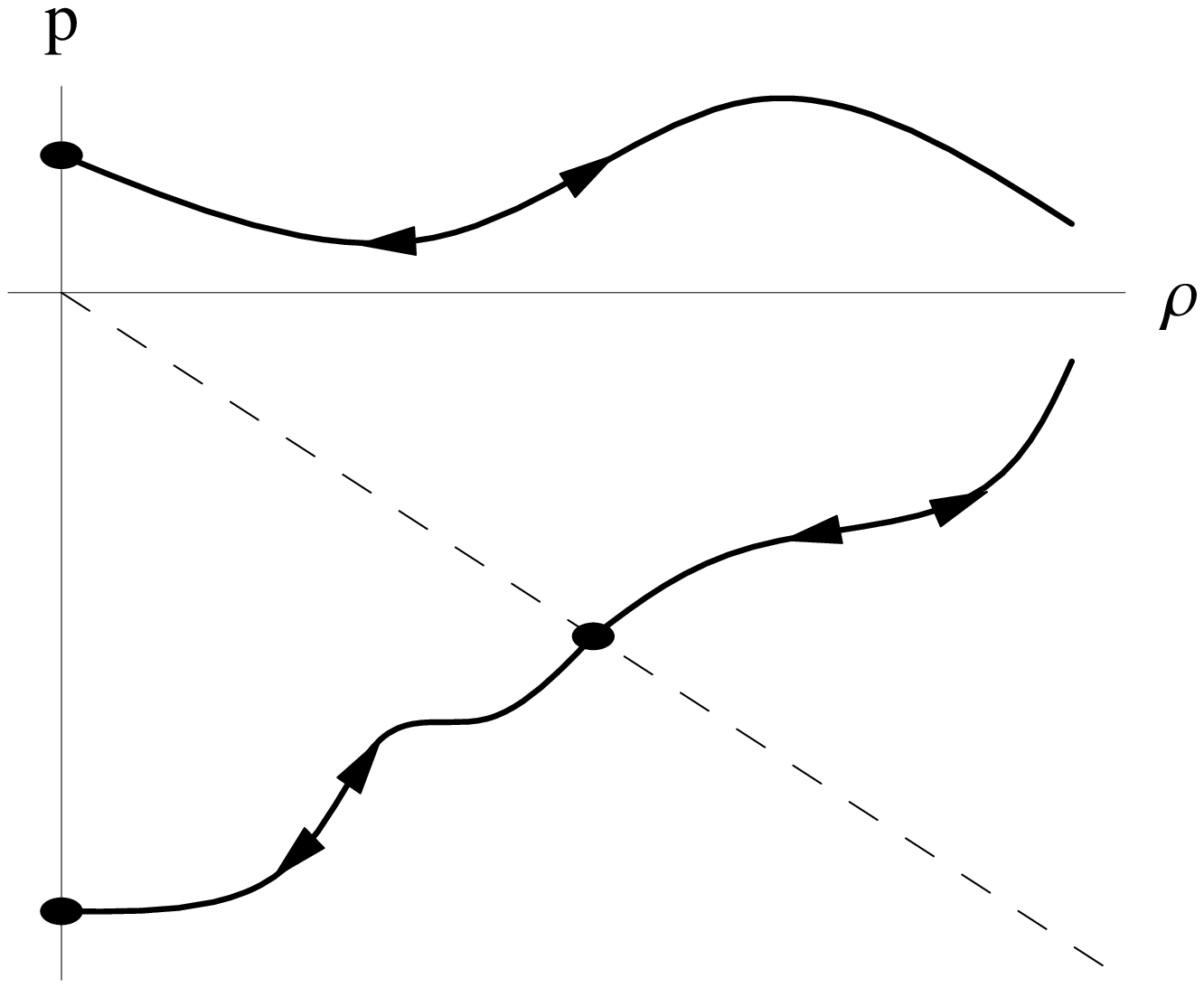}&
\includegraphics[angle=0,width=0.45\textwidth]{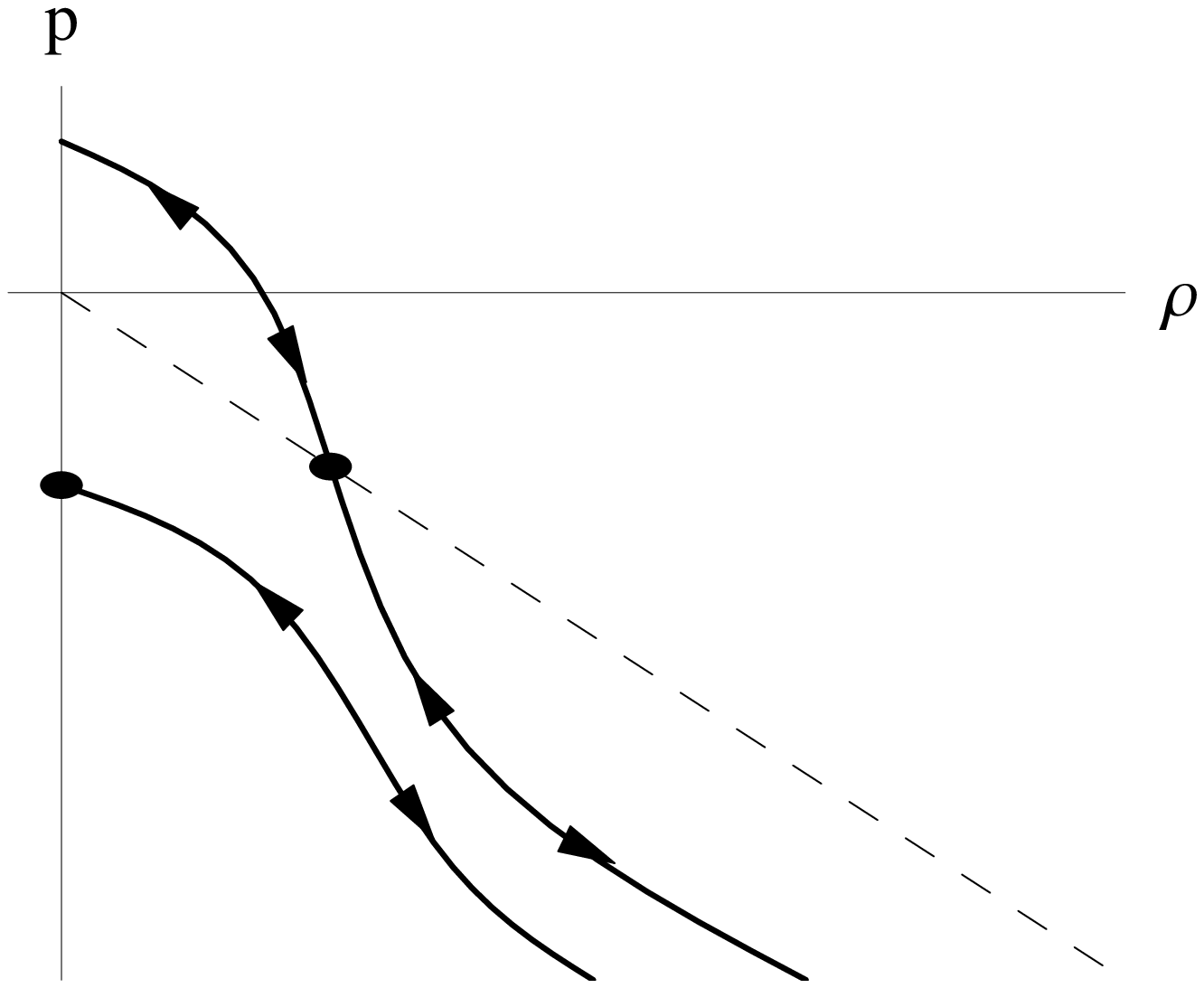}\\
\end{tabular}
  \caption{\label{Gen1}
The evolution of dark energy with an arbitrary equation of state
$p(\rho)$ is shown. The curves $p=p(\rho)$ are the generalization
of a linear equation of state, shown in Figures
(\ref{case1}), (\ref{case2}), (\ref{case3}) and (\ref{case4}).
Near each point of a sufficiently smooth curve $p=p(\rho)$
one can represent the curve by its linear approximation
thus reducing to the results of Section (\ref{Gen}).
}
\end{figure}

\label{Discussion}

In this paper we examined the dynamical evolution of the universe
filled with a dark energy obeying the linear equation of state
(\ref{pDE}). It turns out that this simple linear model  for the
different choices of parameters $\alpha$ and $\rho_0$ has a rich
variety of cosmological dynamics. In dependence on signs of
$1+\alpha$ and $\alpha\rho_0$ and the initial conditions for
$\rho$ and $p$ there can be a set of distinctive types of the
cosmological scenarios: Big Bang, Big Crunch, Big Rip, anti-Big
Rip, solutions with the de Sitter attractor, bouncing solutions, and
their various combinations. In the framework of the linear model
(\ref{pDE}) the analytical solutions of the dynamics of the
universe were obtained. Using the phase plane analysis we gave the
full classification of the solutions in dependence on the
parameters $\alpha$ and $\rho_0$.

We distinguish four main types of the evolution of the
universe filled with dark energy with a linear equation of 
state (\ref{pDE}): 

(i) For parameters $1+\alpha>0$ and $\alpha\rho_0>0$
the universe may contain either  non-phantom or phantom 
energy. For the non-phantom energy there are two types 
of evolution: a) starting from the Big Bang the universe 
approaches to the de Sitter phase during an infinite time; b)
reversed process with respect to the described above 
when the universe starts from a reversed in time the de Sitter phase and evolves to 
the Big Crunch. In the phantom case the universe starts from the 
reversed in time the de Sitter phase then it
riches the bouncing point ($\rho=0$, $p\neq 0$) and after 
bounce the universe approaches to the de Sitter phase.

(ii) For parameters $1+\alpha>0$ and $\alpha\rho_0<0$ the 
universe may contain only non-phantom energy. In this case the only
one type of evolution is possible: the universe expands starting 
from the Big Bang, reaches the bounce point ($\rho=0$, $p\neq 0$) in
finite time and then its expansion changes to the contraction, resulting
in the Big Crunch.

(iii) For parameters $1+\alpha<0$ and $\alpha\rho_0<0$ the universe
may contain either non-phantom or phantom energy. In the case of
phantom universe there are two possible scenarios: a) starting 
from $a_{i}=0$ the universe expands in  
infinite time to the final Big Rip; b) the reversed in time 
process with respect to the described above 
when the universe starts from the anti-Big Rip and evolves to 
final state with $a_f=0$. In the non-phantom case the universe starts from the 
reversed in time the de Sitter phase. Then it
reaches the bouncing point ($\rho=0$, $p\neq 0$) and after 
the bounce the universe approaches the de Sitter phase.

(iv) For parameters $1+\alpha<0$ and $\alpha\rho_0>0$ the 
universe may contain only phantom energy. 
The universe is born in the anti-Big Rip state
and contracts reaching the bouncing point in a finite time. 
After the bouncing the universe begins to expand. 
In a finite time the universe comes to the Big Rip.

The Table \ref{tab1} summarizes the above results for the
evolution of the universe filled with dark energy 
with a generalized equation of state (\ref{pDE}).

It should be stressed that the linearity of a considered dark 
energy equation of state is
not crucial for the general properties of cosmological evolution.
Instead of (\ref{pDE}), one may consider any the rather smooth
curves $p=p(\rho)$ as shown in Fig.~(\ref{Gen1}). It is clear
that the general behavior of the evolution and properties of 
attractors and bounce points do not change in this case because
any sufficiently smooth function $p=p(\rho)$ can be linearized in
the local vicinity of any point. Thus we can reduce the general
problem for $p=p(\rho)$ to the analysis of a linear cosmology
considered in this paper. From the above it follows in particular
that the universe filled with a dark energy with an equation of
state $p_{\rm DE}=p_{\rm DE}(\rho_{\rm DE})$ always approaches the
de Sitter attractor if additionally the physically reasonable
conditions $0<dp_{\rm DE}/d\rho_{\rm DE}<1$ is satisfied. The
first inequality in the last expression means that the considered
dark energy is hydrodynamically stable. While the second
inequality restricts the sound speed to the speed of the light.
A more detailed analysis of the cosmology with an arbitrary
continuous equation of state $p=p(\rho)$ will be presented
elsewhere \cite{BDE_fut}.

Our analysis is limited by the consideration
of the cosmological model of the universe evolution  
filled only with dark energy. Taking into account the 
presence of the ordinary matter and radiation make the 
evolution of the universe much more complicated and is requied
special consideration \cite{BDE_fut}.

\ack

This work was supported in part by the Russian Foundation for
Basic Research grants 02-02-16762-a, 03-02-16436-a and
04-02-16757-a and the Russian Ministry of Science grants
1782.2003.2 and 2063.2003.2.

\end{document}